\documentclass[a4paper,final]{aipproc}
\layoutstyle{6x9}
\title{Selected results and open problems in a semiclassical theory of dense matter}
\author{Vladan \v Celebonovi\'c} {address=
{Inst.of Physics,Pregrevica 118,11080 Zemun-Beograd,Serbia and Montenegro}\\
$vladan@phy.bg.ac.yu$}
\begin{document}
\begin{abstract}
Studies of the behavior of materials under high external pressure
have started in Serbia shortly after the middle of the last
century.The aim of this lecture is to review the theoretical
foundations of this work,present a selection of the results,and
indicate some open problems within the theory.
\end{abstract}
\date{\today}
\maketitle{}
\section{Introduction}
Systematic studies of the behavior of materials under high external
pressure have started near the end of the $XIX$ century.The main "
driving force" behind the rise of interest in this field was a
professor of physics at Harvard,P.W.Bridgman (1882-1961).In 1946.he
was awarded the Nobel prize in physics for {\it "for the invention
of an apparatus to produce extremely high pressures, and for the
discoveries he made therewith in the field of high pressure
physics"}. Presses used by Bridgman were large,expensive to build
and use,and once enclosed in them specimen were no longer visible.
Experimental work in the field was later facilitated with the
discovery of the diamond anvil cell (DAC).The DAC fits into a
hand,it is much cheaper,simple to operate,the upper limit of
pressure is much higher,and the specimen is visible throughout the
experiment.
Modern theoretical interest in the field started in the thrities of the last century when Fermi hinted that increased pressure leads to changes in the atomic structure. The basic aim of these studies is to solve the eigen-problem of a general,non-relativistic many-body Hamiltonian,defined as follows:
\begin{equation}
H=-\frac{\hbar^{2}}{2m}\sum_{i=1}^N\nabla_{i}^{2}+\sum_{i=1}^{N}V(\left|\vec{x_{i}}\right|)+\sum_{i,j=1}^{N}v(\left|\vec{x_{i}}-\vec{x_{j}}\right|)
\end{equation}
According to rigorous quantum mechanics,the energy of the system having a Hamiltonian $H$ is given by
\begin{equation}
    E=\frac{<\Psi|H|\Psi>}{<\Psi|\Psi>}
\end{equation}
All the symbols in these equations have their standard meaning.Sums
in eq.(1) go over the number of particles which,in real physical
situations,is of the order of Avogadro's number.This implies that
these sums can not be calculated exactly. The calculation of such
sums is a serious problem in many-body physics. Once the
eigenproblem of $H$ defined in eq.(1) is somehow solved,the
thermodynamic potentials,phase transition points (or regions) and
the phase diagram in general follow by application of standard
"prescriptions" of statistical mechanics.

This lecture is devoted to a review of a simple theory of dense
matter,whose "`founding fathers"' are two professors of the
University of Belgrade: Pavle Savi\'c and Radivoje Ka\v sanin.As a
consequence of the first letters of their family names it is often
called "the SK theory".It is also often called "the semiclassical
theory"' because it is founded on laws of classical physics and just
a few ideas of atomic physics. The next section contains a review of
the basic physical ideas of their theory,while selected results of
various applications are outlined in the third part.The fourth part
contains a discussion of some of the open possibilities for future
work on this theory,and the lecture ends with the final comments.
\section{The basic ideas}
The development of what later became the SK theory started in 1961
when P.Savi\'c  published a short paper \cite{SAV:61} presenting an
unusual idea: the mean planetary densities,as calculated from the
masses and radii known at the time,could be linked to the mean solar
density by an extremely simple expression:
\begin{equation}
    \rho=\rho_{0}2^{\phi}
\end{equation}

In this expression $\rho_{0}$ denotes the mean solar density,which
at the time was estimated at $4/3$ g$cm^{-3}$. By choosing integral
values of the exponent $\phi$ in the interval $\phi\in(-2,2)$
Savi\'c managed to fit the numerical values of the densities of the
major planets. No hint of any possible physical explanation of this
simple fit was given. In the following 4 years,collaborating with
Radivoje Ka\v sanin, \cite{SK:6265} he managed to develop a theory
of the behavior of materials under high pressure.In later years,it
was nicknamed "the SK theory" after the first letters in their
family names.

The basic physical idea of their theory is simple.SK assumed that
sufficiently high values of external pressure lead to changes of the
electronic structure of atoms and/or molecules. The possibility of
such an interaction was  for the first time hinted by Fermi several
decades before SK. However,according to the avaliable literature,he
simply noted the possibility of such an influence, but did not find
it interesting enough so as to explore it in detail \cite{ES:58}.
The possibility of such an influence has also been invoked around
that time by P.W.Bridgman \cite{ASH:04}. For the special case of a
one-dimensional finite potential well this problem has recently been
discussed by the present author \cite{VL:01}. An analytical
expression for the first pressure derivative of the energy of a
massive particle in such a well has been obtained and discussed in
that paper.

The $SK$ theory is in way similar to the "jelium model" from
classical solid state physics.It represents a material as a uniform
distribution of particles with the mean interparticle separation $a$
defined by
\begin{equation}
    N_{A}(2a)^{3}\rho=A
\end{equation}

In this expression $N_{A}$ denotes Avogadro's number $\rho$ is the
mass density and $A$ is the mean atomic mass of the material.Having
thus introduced $a$,one can define the "`accumulated"' energy per
electron as
\begin{equation}
    E=\frac{e^{2}}{a}
\end{equation}

Logically,one might expect that a relation such as the last one
should contain the ionic charge $Z$It can be shown,(for example
\cite{LEU:84}) that $a$ as defined in eq.(4) is a multiple of the
radius of a Wigner-Seitz cell,which actually contains $Z$.The radius
of a Wigner-Seitz ($WS$) cell is defined by
\begin{equation}
    \frac{A m_{p}}{\rho}=\frac{4}{3}\pi r_{WS}^{3}=\frac{Z}{n}
\end{equation}
It follows from eq.(6) that
\begin{equation}
r_{WS}= (\frac{3}{4\pi})^{1/3}(\frac{A}{\rho N_{A}})^{1/3}=2a(\frac{3}{4\pi})^{1/3}=(\frac{3}{4\pi})^{1/3}(\frac{Z}{n})^{1/3}
\end{equation}

The basic premisses on which the SK theory is based are the following statements \cite{VL:95}
\begin{enumerate}
    \item The density of a material is an increasing function of the pressure to which it is exposed.
    \item With increasing density,every material undergoes a sequence of first order phase transitions.Phases are numbered by an index $i$ and the phase ending at the critical point is denoted as the zeroth phase.

In any given phase (i.e.,for any value of $i$) there exist two limiting values of the density $\rho$ such that
\begin{eqnarray}
    \rho_{i}^{0}\leq\rho_{i}\leq\rho_{i}^{*}\Leftrightarrow
(\frac{1}{\alpha_{i}})\rho_{i}^{*}\leq\rho_{i}\leq\rho_{i}^{*}
\end{eqnarray}
where $\alpha_{i}>1$
\item The maximal densities of two successive phases are related by
\begin{equation}
    \rho_{i+1}^{*}=2\rho_{i}^{*}
\end{equation}
This expression follows directly from eq.(2) by assuming that $\phi_{i+1}-\phi_{i}=1$.

\item It is assumed that
\begin{equation}
    \frac{E_{i}^{*}}{E_{i}^{0}}=\frac{E_{i+1}^{0}}{E_{i}^{*}}
\end{equation}
Some form of a link between the accumulated energies in two
successive phases was needed in order to render the calculations
tractable,and this form was accepted because of its simplicity.After
some algebra,it follows that $\alpha_{i}\alpha_{i+1}=2$ and that
\begin{eqnarray}
\alpha_{i}&= &6/5 , i = 1,3,5,.. \nonumber\\
\alpha_{i}&= &5/3 , i = 2,4,6,..
\end{eqnarray}
\item The final density of the zeroth phase is
\begin{equation}
    \rho_{0}^{*}=\frac{A}{3 \bar{V}}
\end{equation}
which is approximately equal to the critical density in the van der Waals theory. $\bar{V}$ denotes the molar volume of the material at $T=0 K$.In the terminology of the van der Waals theory $\bar{V}=b$.
\item Using assumption 3.,it can be shown that
\begin{equation}
    \frac{A}{\bar{\rho}}=\frac{A}{2}(\frac{1}{\rho_{2}^{0}}+\frac{1}{\rho_{2}^{*}})
\end{equation}

In this expression $\bar{\rho}$ denotes the density at the zero-point,defined as $\bar{\rho}=A/\bar{V}$.
\end{enumerate}

Starting from these premisses,the following set of simple analytical relations can be derived,which are used in all subsequent calculations within this theory.Second order phase transitions can be considered as a special case for which $\bar{V}_{i}^{*}-\bar{V}_{i+1}^{0}\rightarrow0$.

\begin{eqnarray}
\rho_{i}^{*}&=&2^{i}\rho_{0}^{*};   \rho_{i}^{0}=\frac{\rho_{i}^{*}}{\alpha_{i}}\nonumber\\
\bar{V}_{0}^{*}&=&3\bar{V} ; \bar{V}_{i}^{0}=\alpha_{i}\bar{V}_{i}^{*}2^{-i} ;  \bar{V}_{i}^{*}=2^{-i}\bar{V}_{0}^{*} \nonumber\\
r_{0}^{*}&=&(\frac{15}{4 N_{A} 10^{-23}})^{1/3}\bar{V}^{1/3} ; r_{i}^{*}=2^{- i/3}r_{0}^{*} ; r_{i}^{0}=r_{i}^{*} \alpha_{i}^{1/3}
\end{eqnarray}
The following section of this lecture contains a selection of results obtained in various applications of this theory.The algorithm for the calculation of phase transition pressure is relatively short,so it will be presented in detail.The set of equations needed for modelling the internal structure of celestial objects is much longer,so the reader interested in its full details should consult \cite{SK:6265}.
\section{Selected results}
\subsection{Laboratory applications}

The application of the SK theory which is most easily verifiable in
laboratory experiments is the calculation of the value of pressure
on which a first order phase transition occurs in a given
material.This value of pressure can be calculated by considering the
work done by the external pressure in compressing the material,
\begin{equation}
\Delta W=p_{i}^{*}(\bar{V}_{i}^{*}-\bar{V}_{i+1}^{0})= p_{i}^{*}V_{i}^{*}(1-\frac{1}{\alpha_{i}})
\end{equation}

and equating it to the change of the accumulated energy
\begin{equation}
\Delta W=\Delta E = N_{A} (E_{i+1}^{0}-E_{i}^{*})
\end{equation}
Starting from eqs.(11)-(13) it can be shown that the maximal pressure in a phase $i$ of a material is given by
\begin{equation}
p_{i}^{*}\cong1.8077\beta_{i}\bar{V}^{- 4/3} 2^{4 i/3} MBar
\end{equation}
with
\begin{equation}
    \beta_{i}=3\frac{\alpha_{i}^{1/3}-1}{1-1/\alpha_{i}}
\end{equation}
Values of $\alpha$ for various values of the index $i$ are given in eq.(8).
Finally,the value of the external pressure needed to "`provoke"' a first order phase transition from phase $i$ to
the phase $i+1$ in a material is given by
\begin{equation}
p_{tr}=p_{i}^{*}-p_{i}^{0}=p_{i}^{*}(1-2^{-4/3}\frac{\beta_{i-1}}{\beta_{i}})
\end{equation}
This expression gives a simple mathematical procedure for the calculation of the sequence of possible values of a
phase transition pressure in a given material.Those values of pressure at which a first order phase transition is
physically possible are selected by the following criterion:
\begin{equation}
    E_{0}^{*}+E_{I}= E_{i}^{*}
\end{equation}
The symbol $E_{I}$ denotes the ionisation potential,$E_{0}^{*}$ and
$E_{i}^{*}$ can be calculated from eqs.(5) and (12) with $a=10\times
r$ nm. Applying the procedure described above,an analysis of the
applicability of the SK theory to real materials under high pressure
was made \cite{VL:92}. A set of 19 materials for which experimental
data  on phase transitions under high pressure were easily avaliable
was analyzed.The aim was to calculate within the SK theory values of
pressure at which first order phase transitions could be
expected,and then compare the results with the experimental data and
analyze possible causes of the discrepancies.It was shown in
\cite{VL:92} that the relative discrepancies between the measured
values of phase transition pressure and those calculated within the
SK theory are material and pressure dependent.Two basic causes of
the discrepancies were identified:

one is due to that fact that the SK theory takes into account only
the simplest form of the electrostatic potential,while in reality in
atoms and molecules one "deals" with charge distributions;

the second "source of problems" is represented by the fact that
the SK theory neglects the contribution of various non-electrostatic
components to the overall intermolecular potential.This is
expectable for a semiclassical theory,but it clearly increases the
discrepancy between the measured and calculated values of the phase
transition pressure.More details on these two problems are avaliable
in \cite{VL:92}.

Another interesting result of the SK theory concerns the
establishment of the thermal equation of state of solids under
pressure.Namely,the temperature is not explicitly present in the
original formulation of this theory.All its calculations are
performed in the $P-\rho$ plane. However,the temperature can be
introduced in a simple way: by equating the internal energy
densities of a solid,as expressed within the SK theory and in
standard solid state physics.

The internal energy per unit volume of a solid is,in the SK theory,given by

\begin{equation}
E=2e^{2}Z(\frac{N_{A}\rho}{A})^{4/3}
\end{equation}

and in standard solid state physics,it is given by
\begin{equation}
E=\frac{\pi^{2}(k_{B}T)^{4}}{10(\hbar\bar{u})^{3}}
\end{equation}

The symbol $\bar{u}$ in the last expression denotes the mean value
of the velocity of sound,and all the other symbols have their
standard meanings.

Details are avaliable in \cite{VL:91} and the final result is that the equation of state of a solid in the $T-\rho$
plane has the following form:
\begin{equation}
    T \left[K\right]=1.4217\times 10^{5} (\frac{\rho}{A})^{7/12} (\frac{m_{e}}{M})^{3/8} Z^{7/8}
\end{equation}
The symbol $m_{e}$ denotes the bare electron mass,$M$ is the ionic mass,$A$ denotes the mass number of the material and $Z$ is the charge of the ions. A result of an important astronomical application of this equation of state is presented in the following subsection of this contribution.

Last,but certainly not the least,we come to the problem of hydrogen
under high pressure,and its possible metallization. The SK theory
has been applied to this problem \cite{VL:89},and the result
obtained at the time was encouraging. Metallisation was predicted to
occur at P=$300$ GPa,which agreed with predictions by various other
theoretical methods. Theoretical predictions of the metallisation
pressure of hydrogen have been of this order of magnitude for the
last 70 years \cite{WH:35}.

The experimental situation in work on hydrogen under high pressure
is much less well defined.There have been claims at the beginning of
the nineties that metallisation occurs at a pressure $P=150$ GPa.A
short time after these results were shown to be incorrect and due to
a chemical reaction of ruby with hydrogen \cite{MHH:92}. Static
experiments performed at values of pressure $P\leq342$ GPa have not
shown signs of metallization of hydrogen \cite{NAR:98},so the
problem seems (at present)to be completely open.The literature on
this topic is huge,but for examples of interesting recent papers see
\cite{NELL:01} and \cite{ASH:04}.One of these papers
,\cite{NELL:01},invokes the influence of disorder as a possible
reason why metallization was not observed at the theoretically
predicted value of the pressure.

\subsection{Applications in astronomy}

The theory we are discussing in this lecture can be applied to
modellization of the internal structure of celestial objects.To be
precise,the interest of Savi\'c and Ka\v sanin in this problem was
the initial "grain of salt" which led to its development.As the
calculations in this theory do not contain any mention of internal
energy generation in celestial objects (i.e.,nuclear reactions),it
is inapplicable to work on stellar structure.

The complete calculational scheme for astronomical applications of the SK theory is avaliable in \cite{SK:6265}.
Unfortunately,due to various reasons,it has not been reformulated in a modern way. The only input data needed for
making the model of the internal structure of a celestial object within this theory are the mass and the radius of
the object.

Starting from these data,it gives the number and thickness of layers
which exist in the interior of the object,the distribution of the
values $P,\rho,T$  with depth under the surface,the strength of the
magnetic field and the interval of the physically allowed values of
the speed of rotation.The theory also gives as a result the mean
atomic(or molecular) mass of the chemical mixture that the object
under study is made of.

The first celestial body to be modelled was logically the Earth and the main characteristics of this model are shown
in the following table.
\medskip
\begin{table}[ht]
\begin{tabular}{lrrrr}
\hline
depth (km) & 0 - 39 & 39 - 2900 & 2900 - 4980 & 4980 - 6371 \\
$\rho_{max}$[kg/m$^3$] & 3000 & 6000 & 12000  & 19740 \\
$P_{max}$[GPa] & 25 & 129 & 289 & 370 \\
$T_{max}$[K] & 1300 & 2700 & 4100  & 7000 \\
\hline
\end{tabular}
\caption{The interior of the Earth according to SK}
\label{tab:1}
\end{table}

The mean mass number of the material which makes up the
Earth is $A=26.56$.
Taking into account the simplicity of the theory,
this model is in remarkable
agreement with modern knowledge.For a discussion
of the temperature in the interior of the Earth,see for example \cite{DH:95}.

Note also that a current model of the Earth's crust called CRUST 5.1,\cite{USGS} gives around $70$ km as the maximal value.Here in Holland,the thickness of the Earth's crust accoring to this model is around 35 km,which is in excellent agreement with the value calculated within SK.

Apart the Earth,the theory was applied to all the other planets except Saturn and Pluto,the Moon,the Galileian satellites of Jupiter,the satellites of Uranus,Neptune's satellite Triton and the asteroids 1 Ceres and 10 Hygiea.

The results are scattered in the literature,but a "safe" general comment is that the agreement with the consequences of observations and with theoretical work of other authors is good \cite{VL:95}.

The following table contains the values of the mean atomic masses of various objects in the Solar System,calculated accoring to the SK theory.
\medskip
\begin{table}[ht]
\begin{tabular}{lrrrr}
\hline
object & A & satellite & A \\
\hline
Sun & 1.4 & Moon & 71 \\
Mercury & 113 & J1 & 70 \\
Venus & 28.12 & J2 & 71 \\
Earth & 26.56 & J3 & 18 \\
Mars & 69 & J4 & 19 \\
1 Ceres & 96 & U1 & 38 \\
Jupiter & 1.55 & U2&  43\\
Saturn & / & U3 & 44\\
Uranus & 6.5& U4 & 32\\
Neptune & 7.26& U5 & 32\\
Pluto &  / & Triton & 67\\
\hline
\end{tabular}
\caption{The composition of the Solar System according to SK}
\label{tab:2}
\end{table}

It can easily be seen from the preceding table that our planetary
system is far from being chemically homogenous;at first sight,the
well known division on the terrestrial and jovian planets is clearly
visible. These differences are obviously a "remnant" of various
transport and mixing processes which have been active in the
formation epochs of the planetary system.Note also that similar
differences are visible in the satellite systems which were modeled.

Various  conclusions can be drawn from data in Table 2.
For example,asteroid 1 Ceres is currently orbiting the Sun between the orbits of Mars and Jupiter.However,by its
chemical composition it is similar to the planet Mercury \cite{VL:95}.
As chemically similar bodies are expected to have been formed close
to each other,this similarity implies that {\it " once upon a time"
} Ceres and Mercury originated in the same region of the
protoplanetary system,but that their orbits later diverged.The
physical process(es) which have led to this diverging of their
orbits can be a subject of further studies.

Concerning asteroids,using the value of $A$ calculated for 1 Ceres,the mass of the asteroid 10 Hygiea was calculated,
and the result turned out to be in excellent agreement with the result known in celestical mechanics \cite{VL:88}.

Two cases of applications of the SK theory are especially worth
mentioning.The composition  of the Galileian satellites was
determined within this theory in 1987. In 1996.some results of this
calculation were confirmed by measurements from the Galileo space
probe.Details are given in \cite{VL:98} which contains the reference
to the original publication of 1987.

Another interesting application of this theory concerns Neptune's
satellite Triton \cite{VL:86}.It was shown in that paper that Triton
has a composition similar to Mars,which was interpreted as implying
that Triton is a captured body,in perfect agreement with earlier
work based purely on celestial mechanics.This conclusion is of
interest for cosmogony,as it can be regarded as an independent proof
that collisions were important in the early phases of existence of
our planetary system.For a recent example of observational work on
the structure of a protoplanetary cloud see \cite{SUB:04}.

\section{Some possibilities for future work}
In this lecture we have so far reviewed the basic ideas and a number of results of applications of the SK theory.
 However,in spite of successes,there are open possibilities for future work related to the theory we are discussing,and
 this section is devoted to an outline of these topics.

Paper \cite{SAV:61} was devoted to a possible relation tying the
mean planetary densities (derived from their masses and radii) with
the mean solar density.Eq.(1) was the result,and it was applied in
various calculations in its original form ever since.The important
detail in that paper was not only the simple form of eq.(1),but also
the fact that the exponent $\phi$ in eq.(1) had {\it integral}
values.However,as nearly 40 years have elapsed since
\cite{SAV:61},it seemed appropriate to undertake a verification of
eq.(1) but with modern data,and on a broader set of objects. Such a
check has recently been performed \cite{VL:04} and the results are
shown in table 3.

\medskip
\begin{table}[ht]
\begin{tabular}{lrrrr}
\hline
object & $\rho_{mean}$ & $\phi$ & $\rho_{calc}$\\
\hline
Sun & 1408 & 0 & 1408 \\
Mercury & 5427 & 2 & 5632 \\
Venus & 5243 & 2 & 5632 \\
Earth & 5515 & 2 & 5632 \\
Moon & 3350 & 5/4 & 3349\\
Mars & 3933& 3/2 & 3982 \\
Phobos& 1900 & 2/5 & 1858 \\
Deimos & 1750 & 1/3 & 1774\\
Jupiter & 1326 & - 1/10 & 1314\\
JI & 3530 & 7/5 & 3716 \\
JII & 3010 & 1 & 2816\\
JIII & 1940 & 1/2 & 1991\\
JIV & 1840 & 2/5 & 1858\\
Saturn & 687 & - 1 & 704\\
Titan & 1881 & 2/5 & 1858\\
Uranus & 1270& -1/7 & 1275\\
Ariel & 1670 & 1/4 & 1674\\
Neptune & 1638 & 1/4 & 1674\\
Triton & 2050& 1/2& 1991\\
Pluto & 1750 & 1/3 & 1774\\
Charon & 2000 & 1/2 & 1991\\
\hline
\end{tabular}
\caption{Modern values of the exponent $\phi$}
\label{tab:3}
\end{table}

Modern data on densities of 22 planets and their major satellites
were used.It was shown that eq.(1) is still valid,but that the
exponent $\phi$ can also take {\it non-integral} values. The
immediate question is of course the interpretation of these
non-integral values.Namely,the integral values of $\phi$ which
appeared in \cite{SAV:61} were interpreted there and in later
publications as a qualitative analogy with atomic quantisation,
implying that high external pressure leads to changes in the atomic
structure.The problem is open,and one of the possibilities would be
to try to cautiously pursue the analogies between the atomic
structure and the structure of the planetary system.

Another "`open"' astronomical application of the SK theory is the
calculation of the angular speed of rotation of a celestial
object.Instead of proposing an algorithm for this calculation and
giving a unique number at the end,this theory gives a physically
allowed interval in which the speed of rotation can be.

It was applied to the Earth and several other planets. The results
are very promising,but it would be useful to reduce this allowed
interval to a unique number.

A  possibility for future work concerns laboratory applications of
the SK theory.We have discussed to some extent the applicability of
this theory to the calculation of phase transition pressure in
various solid materials,and to the establishment of a thermal
equation of state of solids. The discrepancies between the predicted
values of phase transition pressure and experimental results are
material dependent. Work in this direction could in the future
advance along several different lines:

At first it would be useful to apply the theory to more materials
for which first order phase transition presssures are known.In this
way one would obtain a more precise empirical estimate of the
systematic trends (if there are any) in the discrepancies between
the calculated and measured values of the phase transition pressure.

On the purely theoretical side,in would be necessary to refine the theory with
the inclusion of the contribution of more
components to the interparticle potential energy.

An even more interesting problem would be to link the SK theory to
the well established theoretical framework of statistical
physics.Although the SK theory has given good results in a range of
applications,its unusal formulation hinders its wider spread in the
research community.
A preliminary attempt in that direction has recently been performed
in \cite{VL:04},where is was shown how the parameters appearing in
the SK theory can be connected with those of the Landau theory. It
would be interesting trying to reformulate the SK theory along the
lines of present day theories of quantum phase transitions (for
example,\cite{VO:03}).

A note concerning the equation of state of solids under pressure in
$T-\rho$ plane:

This calculation,described in detail in \cite{VL:91} gives physically acceptable numerical
values,which is good.However,the calculation invokes the speed of sound in a solid,and (as well known)
this invokes the knowledge of $\partial P/\partial\rho$ - that is the equation of state.

In the calculation which has led to the proposed form of thermal
equation of state,an approximation called the "`Bohm-Staver"'
formula was used \cite{AM:76}.It is planned in the future to repeat
this calculation,but using some more refined form of the equation of
state.

\section{Final Comments}

This lecture was an attempt to present a "balanced" review of  a
theory of dense matter proposed a little less than half a century
ago by P.Savi\'c and R.Ka\v sanin. The adjective "balanced" simply
means that the basic ideas,successful applications,but also problems
for future work were discussed.In the past,work within the SK theory
has been more oriented towards various applications and to a lesser
extent towards its modernization.In future work, efforts will
concentrate more on the refinement of  this theory and extension of
its range of applicability. It is hoped that as result of this
workshop some "joint venture" in that direction will be started.

\section{Acknowledgements}
The author is grateful to Prof.Wim van Saarloos,director of the Lorentz Center for making the initial proposal for the organization of this workshop,and to all in the Lorentz Center whose work helped that the workshop runs so smoothly and succesively.The preparation of this contribution was financed by the Ministry of Science and Environment of Serbia within its project 1231.
.

\end{document}